\documentstyle[aps,12pt]{revtex}
\begin{document}
\title{On the relationship between the angular deficit and the internal
structure
of straight cosmic strings.}
\author{F. Dahia\thanks{%
E-mail: fdahia@fisica.ufpb.br}}
\address{Departamento de F\'{\i}sica, Universidade Federal da
Para\'{\i}ba,\\
Cx. Postal 5008 58059-970,\\
Jo\~{a}o Pessoa, Pb, Brazil.}
\maketitle

\begin{abstract}
We give a brief discussion on the limitations involving the expression $\mu
=\Delta \phi /8\pi $ $(G=c=1),$ which relates the string linear energy
density $\mu $ to the conical deficit angle $\Delta \phi $. Then, we
establish a new equation between the angular deficit and a combination of
the components of the string stress-energy tensor which shows that the
angular deficit is determined not only by the amount of proper matter of the
string but also, in a Newtonian sense, by its internal gravitational field.
\end{abstract}

\vspace{1.0in}Cosmic strings are topological defects\cite{vilenkin} which
produce very interesting gravitational effects. We find in the literature a
great number of articles dealing with this subject (see Ref. \cite{vilenkin}
for a extensive reference list). In this paper we want to focus our
attention on a specific aspect of cosmic strings. It is well known that the
asymptotic space-time generated by an infinite straight cosmic string has no
curvature, but has an angular deficit \cite{vilenkin,garfinkel}. It is
argued in many works that this angular deficit can be connected with a
physical property of the string, namely, its ``linear energy density'' by
the expression $\mu =\bigtriangleup \phi /8\pi $ \cite
{vilenkin,garfinkel,hiscock,gott}. Thus, it would be possible to infer the
cosmic string linear energy density measuring the space-time angular
deficit. We would like to investigate this question more closely.

The cosmic string in a curved space-time is a special solution of the
Einstein equations coupled with scalar and gauge fields in which the
stress-energy tensor is concentrated around a ``line'' of false vacuum. The
non-gravitational fields are described by the Lagrangian \cite{garfinkel}:
\begin{eqnarray}
{\cal L} &=&-\frac 12\nabla ^\nu R\nabla _\nu R-\frac 12R^2(\nabla _\nu \psi
+eA_\nu )(\nabla ^\nu \psi +eA^\nu )  \nonumber  \label{lagrangian} \\
&&-\alpha \left( R^2-\eta ^2\right) ^2-\frac 1{16\pi }F_{\nu \kappa }F^{\nu
\kappa },  \label{lagrangian}
\end{eqnarray}
where $A_\nu $ is a vector field, $\Phi \equiv R\,e^{i\psi }$ is a complex
scalar field, $\nabla _\nu $ is the covariant derivative with respect to the
space-time metric, $F_{\nu \kappa }\equiv \nabla _\nu A_\kappa -\nabla
_\kappa A_\nu ,$ $\alpha ,$ $\eta $ and $e$ are constants. (Note that $\nu
,\kappa =0,1,2,3,$ and $\hbar =1)$. Actually, there are alternatives
proposals for string models \cite{vilenkin}, but here we are considering the
Abelian-Higgs model (\ref{lagrangian}), which is the simplest one.

The exact string solution for the Lagrangian (\ref{lagrangian}) is not known
but it is assumed that the fields have the form:
\begin{eqnarray}
R &=&R(r),  \label{ansatz-R} \\
\psi &=&\phi ,  \label{ansatz-psi} \\
A_\nu &=&\frac 1e\left[ P(r)-1\right] \nabla _\nu \phi .  \label{ansatz-A}
\end{eqnarray}
It is expected that $R\rightarrow \eta $ and $P\rightarrow 1$ as $%
r\rightarrow \infty .$ This asymptotic behavior of the fields ensures that
the stress-energy tensor drops to zero far from the string axis \cite
{garfinkel}. Other relevant property, which comes from
(\ref{ansatz-R}),(\ref
{ansatz-psi}) and (\ref{ansatz-A}) is that the associated stress-energy
tensor is static, has cylindrical symmetry and it is boost-invariant along
the axis direction. It is reasonable to assume the same symmetries for the
metric. The most general metric with these symmetries can be put in the
form:
\begin{equation}
ds^2=e^{2V(r)}(-dt^2+dz^2)+dr^2+U^2(r)d\phi ^2,  \label{metric}
\end{equation}
where $-\infty <t,z<\infty ,$ $0<r<\infty $ and $0<\phi <2\pi .$ The
regularity condition on the string axis imposes that $\frac Ur\rightarrow 1$
as $r\rightarrow 0$ and that $V(r)$ must be a well behaved function on the
axis.

According to Garfinkel \cite{garfinkel} and other authors \cite
{vilenkin,linet,frolov}, the linear energy density $\mu $ of the string is
defined as the integral of the energy density $\rho =-T_t^t,$ which is
measured by the observers associated with the vector field $%
e^{-V(r)}\partial _t,$ over the transversal section $t=const$ and
$z=const:$%
\begin{equation}
\mu \equiv -\int \int T_t^tU(r)drd\phi .  \label{mi}
\end{equation}

With some additional assumptions, Garfinkel \cite{garfinkel} has shown that
the metric (\ref{metric}) generated by the cosmic string is asymptotically a
conical space-time with $UV^{\prime }\rightarrow 0$ as $r\rightarrow \infty
. $ Taking into account these results and using the Einstein equation, $8\pi
UT_t^t=U^{\prime \prime }$ $+(UV^{\prime })^{\prime }+UV^{\prime \,2},$ one
can establish, from definition (\ref{mi}), a relation between $\mu $ and the
angular deficit $\bigtriangleup \phi \equiv 2\pi [1-U^{\prime }(\infty )]$
of the conical space:
\begin{equation}
\mu =\frac{\bigtriangleup \phi }{8\pi }-\frac 14\int UV^{\prime \,2}dr.
\end{equation}

When the energy of the false vacuum is null ($\eta =0)$, the metric reduces
to the Minkowski metric. Then, Garfinkel \cite{garfinkel} argued that for
small $\eta ^2$ ( the estimated value for grand unified strings is $\eta
^2<10^{-4}),$ the metrics components will have little deviation from the
flat metric values. Based on this approximation, he estimated the order of
the correction for $U$ and $V$ and, then, showed that $\mu $ is of the order
of $\eta ^2$ but the integral term is of the order of $\eta ^4.$ Thus, to a
good approximation:
\begin{equation}
\bigtriangleup \phi \simeq 8\pi \mu .  \label{deficit}
\end{equation}

The complicated form of the full Einstein-gauge-scalar equations for the
strings does not recommend that we look for the exact string solution. Thus,
we have to restrict ourselves to approximated relations such as equation (%
\ref{deficit}). To get further insight about the exact form of the metric in
the string solutions, some authors propose to use simple models to describe
cosmic strings.

It is estimated that the transverse dimensions of the string is very small
compared with its length. Then, it may be appropriate treating the string as
an energy distribution concentrated around an axis in such a way that the
string can be confined in a thin cylinder. In addition, it would be
reasonable to assume that the stress-energy tensor has all the symmetries
mentioned previously. Hiscock \cite{hiscock} and Gott \cite{gott} studied
the simplest model with these properties. In their model the stress-energy
tensor has an uniform energy density $\rho $ and it is given by
\begin{equation}
T_t^t=T_z^z=-\frac{\gamma ^2}{8\pi \ell ^2}=const,\qquad (r\leq \ell ),
\end{equation}
with all other components equal to zero. The parameter $\ell $ is the
cylinder radius. For $r>\ell $, the stress-energy vanishes, since we are
considering that the string is surrounded by vacuum.

The space-time generated by this energy distribution can be determined. We
can distinguish between two regions, inside and outside the cylinder, to
which correspond respectively an interior and an exterior metric that can be
smoothly joined :
\begin{eqnarray}
ds^2 &=&-dt^2+dr^{\prime \,2}+\left( \ell /\gamma \right) ^2\sin ^2\left(
\gamma r^{\prime }/\ell \right) d\phi ^2+dz^2,\qquad r^{\prime }\leq \ell
\label{interior} \\
ds^2 &=&-dt^2+dr^2+\lambda ^2r^2d\phi ^2+dz^2,{\qquad }r>r(\ell),
\end{eqnarray}
with $\lambda =\cos \gamma$ and $r(\ell)=(\ell / \gamma)\tan \gamma$. These latter relations come from the continuous junction condition through the cylinder
surface $r^{\prime }=\ell$ (or $r=r(\ell)$) , which demands the metric and extrinsic
curvature of the cylinder surface to be both continuous.

Regarding the definition of the linear energy density, it can be shown that
the angular deficit is
\begin{equation}
\Delta \phi =8\pi \mu ,  \label{exact}
\end{equation}
which coincides with the approximate result (\ref{deficit}), but now holds
for all orders in $\mu .$ Note that this equation is independent of the
cylinder radius. Then, taking the limit $\ell \rightarrow 0,$ the
distribution can be idealized as a line source with a well defined linear
energy density which produces a conical space-time with an angular deficit
proportional to $\mu .$

The generality of this result seems to be questionable, since it was derived
from a string whose stress-energy tensor has a very simple form. But the
strings models can be enlarged as was demonstrated by Linet \cite{linet},
who showed that for models with non-uniform energy the relation
(\ref{exact}%
) is also satisfied. Then, we could be led to suppose that this is a general
characteristic, i.e., that the angular deficit of a conical space-time
generated by a line source is given by (\ref{exact}).

However this conclusion is not correct as was shown by Geroch and Traschen
\cite{geroch}, who considered another string model for which the interior
metric corresponds to (\ref{interior}) multiplied by a conformal factor:
\begin{equation}
ds^2=\exp [2\beta f(r^{\prime }/\ell )]\,[-dt^2+dr^{\prime \,2}+(\ell
/\gamma )^2\sin ^2(\gamma r^{\prime }/\ell )d\phi ^2+dz^2],\text{\qquad }%
(r^{\prime }\leq \ell )
\end{equation}
where $f$ is a smooth non-negative function with compact support inside the
interval $\left[ \frac 12,1\right] .$ The properties of the function $f$ at
$%
r^{\prime }=\ell $ ensure a continuous junction with the exterior conical
metric. Moreover, the associated stress-energy tensor has all desirable
symmetries. Hence, the metric above is, in fact, a possible interior
solution. Now, calculating the linear energy density we obtain
\begin{equation}
\mu =\frac{\bigtriangleup \phi }{8\pi }-\frac{\beta ^2}{4\gamma }\int_{\frac
12}^1\sin \left( \gamma x\right) \left[ f^{\,\prime }(x)\right] ^2dx.
\end{equation}
We can choose $\beta $ in such a way that $\mu $ is positive. Then, taking
the limit $\ell \rightarrow 0,$ we get a line source for the conical metric
with $\mu $ less than the angular deficit. Thus, according to Geroch and
Traschen \cite{geroch}: ``the procedure of taking the limit of a family of
well-behaved sources does not in general yield an unique relationship
between the mass per unit length and deficit angle''. This result was
investigated and confirmed by other authors who have shown that, except in
some special case when the constants of the Lagrangian $e$ and $\alpha $
satisfy a particular tuning \cite{linet1}, there is no simple relation
between $\mu $ and $\Delta \phi $ in the thin limit of cosmic strings \cite
{futamase,hindmarsh}.

So, we have to specify stronger conditions for the string models, imposing a
great number of constraints to the internal distribution, if we want to
ensure the validity of relation (\ref{exact}), see Refs.\cite
{linet1,futamase,israel}. Another possibility is to restrict the space-time
to a special class of manifolds for which the distributional formalism can
be used to treat conical singularities \cite{dahia}.

We would like to approach this problem in another way. Instead of trying to
determine the conditions to be satisfied by the internal model in order to
preserve relation (\ref{exact}), we admit that (\ref{exact}) may be
incomplete and ask for the existence of another physical attribute of the
source that could be unambiguously associated with the angular deficit.

Firstly, consider the metric in the Gaussian form:
\begin{equation}
ds^2=dr^2+g_{ij}dx^idx^j,\quad \quad i,j=1,2,3
\end{equation}

The Einstein equations in this coordinate system can be written in the
following way:
\begin{eqnarray}
\frac 1{\sqrt{-g}}\frac \partial {\partial r}\left[ \sqrt{-g}\left(
K_j^i-\delta _j^iK\right) \right] +\left( ^{(3)}R_j^i-\delta _j^i\,^{\left(
3\right) }R\right) &=&-8\pi \left( T_j^i+\delta _j^iT_r^r\right) , \\
K,_i-\,^{(3)}\nabla _jK_i^j &=&-8\pi T_i^r, \\
K_j^iK_i^j-K^2-\,^{\left( 3\right) }R &=&-16\pi T_r^r,
\end{eqnarray}
where $K_j^i=\frac 12g^{ik}g_{kj,r}$ is the extrinsic curvature of the $%
r=const$ hypersurfaces, $K$ is the trace of $K_j^i,$ and $^{\left( 3\right)
}R_j^i$ is the Ricci tensor of the hypersurface $r=const.$

We consider an internal model whose associated stress-energy tensor has all
characteristics mentioned previously. So, we will admit that the interior
metric can be put in the form (\ref{metric}) with the same regularity
conditions. The extrinsic curvature of the $r=const$ hypersurface is easily
calculated: $K_t^t=K_z^z=V^{\prime }$ and $K_\phi ^\phi =U^{\prime }/U,$
where prime denotes derivative with respect to $r.$ For this metric, the
Einstein equations reduce to the form:
\begin{eqnarray}
\frac \partial {\partial r}\left[ \sqrt{-g}\left( K_j^i-\delta _j^iK\right)
\right] &=&-8\pi \sqrt{-g}\left( T_j^i+\delta _j^iT_r^r\right) ,
\label{eq1}
\\
0 &=&T_i^r, \\
K_j^iK_i^j-K^2 &=&-16\pi T_r^r,  \label{vinc}
\end{eqnarray}

Integrating equation (\ref{eq1}) over the transversal section of the string,
the surface $t=const$ and $z=const,$ we obtain:
\begin{equation}
\left. \sqrt{-g}\left( K_j^i-\delta _j^iK\right) \right| _0^\ell
=-4\int_0^{2\pi }\int_0^\ell \sqrt{-g}\left( T_j^i+\delta _j^iT_r^r\right)
drd\phi .  \label{int-eq1}
\end{equation}

We are looking for interior metrics that can be matched smoothly with a
conical metric. On the cylindrical hypersurface $r=\ell ,$ the continuity
conditions for the metric and extrinsic curvature requires that:
\begin{eqnarray}
V(\ell ) &=&0; \\
V^{\,\prime }(\ell ) &=&0;\quad U^{\prime }(\ell )=\lambda .
\end{eqnarray}

Then, from these conditions and equation (\ref{int-eq1}), we obtain
\begin{equation}
\left( \frac{\Delta \phi }{8\pi }-\chi \right) \left( \delta
_{\;j}^i-\,\delta _{\;\phi }^i\,\delta _{\;j}^\phi \right) =-\int_0^{2\pi
}\int_0^\ell \sqrt{-g}\left( T_j^i+\delta _j^iT_r^r\right) drd\phi
\label{new}
\end{equation}
where $\chi =\left( 1-e^{2V\left( 0\right) }\right) /4$.

Now in order to write $\chi $ in terms of the physical attributes of the
cosmic string let us consider constraint equation (\ref{vinc}) and the
equation of energy-momentum conservation $\nabla _\kappa T_{\;\nu }^\kappa
=0:$%
\begin{eqnarray}
V^{\prime \;2}+2V^{\prime }\frac{U^{\prime }}U &=&8\pi p_r \\
p_r^{\prime }+2\left( p_r^{}+\rho \right) V^{\prime }+\left( p_r-p_\varphi
\right) \frac{U^{\prime }}U &=&0
\end{eqnarray}
From these equation we can show that
\begin{equation}
V\left( 0\right) =\int_0^\ell \left[ \frac{p_r^{\prime }-\sqrt{p_r^{\prime
\;2}-8\pi p_r\left( p_r-p_\varphi \right) \left( 3p_r+4\rho +p_\varphi
\right) }}{\left( 3p_r+4\rho +p_\varphi \right) }\right] dr
\end{equation}
where $\rho =-T_t^t$, $p_r=T_r^r$ and $p_\phi =T_\phi ^\phi $ are the
volumetric energy density, the radial pressure and the hoop pressure
respectively, measured by observers associated to the field $e^{-V}\partial
_t.$

For $i,j=t$, we have
\begin{equation}
\frac{\Delta \phi }{8\pi }=\int_0^{2\pi }\int_0^\ell \sqrt{-g}\left( \rho
-p_r\right) drd\phi +\chi .  \label{main}
\end{equation}

For $i,j=\phi $, we have the well known constraint for cosmic strings \cite
{hindmarsh}:
\begin{equation}
\int_0^{2\pi }\int_0^\ell \sqrt{-g}\left( p_r+p_\phi ^{}\right) drd\phi =0.
\end{equation}
To avoid misunderstanding it is worthy to emphasize here that the above
equation does not imply that the integral of each component $T_r^r$ and $%
T_\phi ^\phi ,$ taken separately, also vanishes. Thus, the additional terms
that appear in (\ref{main}) cannot be neglected. However, the integrals of
the ``Cartesian'' components $T_x^x$ and $T_y^y$ (where $x=r\cos \phi $ and
$%
y=r\sin \phi $) are both null \cite{moss,peter}$.$ At first sight this could
seem contradictory, since the new terms in (\ref{main}) depend on the transversal components
of pressure whose integral, as we said before, vanish in Cartesian
coordinates. This point can be clarified by noting that the expression (\ref
{main}) is not covariant, since it holds only in coordinates in which the
Einstein equations assume the form (\ref{eq1}), which can be directly
integrated. In Cartesian coordinates the correspondent equations are not so
simple, except in the linearized approximation. In this regime, as was shown
in \cite{peter}, for Abelian-Higgs cosmic strings, the deficit angular
depends only on the linear energy density according to (\ref{mi}). But this
result is exactly what we obtain from (\ref{main}) in the linear
approximation. Indeed, we can verify that, in this limit, $\chi \simeq
\int_0^{2\pi }\int_0^\ell \sqrt{-g}p_rdrd\phi $, and hence the formula (\ref
{main}) reproduces equation (\ref{mi}).

Equation (\ref{main}) is an exact formula which generalizes (\ref{exact}).
We can notice important differences by comparing them. In (\ref{main}) the
integration is done taking the volume $\surd -g$ of the whole space-time,
while definition (\ref{mi}) uses the area element $\surd ^{\left( 2\right)
}g $ induced on the surface $t=const$ and $z=const.$ This replacement has a
physical meaning which we shall try to interpret now. The factor $\surd
^{\;\left( 2\right) }g$ gives only the correction for the area element $%
da=\surd ^{\;\left( 2\right) }gdrd\phi $ due to the curvature of the surface
$t=const$ and $z=const.$ On the other hand, $\surd -g=\surd -g_{tt}^{}\surd
^{\;\left( 3\right) }g$ contains two distinct contributions: one due to the
spacial curvature incorporated by $dV=\surd ^{\;\left( 3\right) }gdrd\phi
dz, $ which is the proper volume element of the three-space sliced by the
observers $e^{-V}\partial _t$; and another contribution corresponding to the
term $\surd -g_{tt}$. As we know, in the weak field approximation, this
component $g_{tt}$ is related to the Newtonian gravitational potential $\psi
$ ($g_{tt}\simeq 1+2\psi )$. This is an interesting result because,
expanding (\ref{main}) in this approximation, we obtain a term proportional
to $\frac 12\rho \psi $, which is the gravitational energy of the Newtonian
field $\psi $ generated by ordinary (non-relativistic) matter distribution.
Strictly speaking, in the string case, there is no Newtonian limit properly,
since the energy distribution is ultra-relativistic, once the pressure and
the energy are of the same order \cite{garfinkel}. Because of this and the
fact that the state equation $p_z=-\rho $ holds for the cosmic string, the
analogous of the Newtonian potential satisfies $\nabla ^2\psi =4\pi \left(
p_r+p_\phi \right) ,$ instead of the common result $\nabla ^2\psi =-4\pi
\rho .$ Hence, the term $\frac 12\rho \psi $ is not exactly the Newtonian
gravitational energy density. However, it shows that the gravitational
potential also contributes to the angular deficit, not only the amount of
matter, as it would be the case if $g_{tt}$ were not present in the formula
(%
\ref{main}). Thus, in this sense, through the factor $\surd -g$ in (\ref
{main}), the gravitational field contributes itself to the angular deficit.

Besides this, the presence of the additional term related to radial pressure
highlights the influence of the internal gravitational field on the angular
deficit, since the static configuration is reached only after the
equilibrium between the radial pressure and the gravitational force is
attained. Finally, we have the term $\chi $ which is associated to the
gravitational red shift between the axis and the hypersurface at $\ell .$
Thus, equation (\ref{main}), which reminds us of Tolman's integrals \cite
{tolman}, suggests that the gravitational field (in a Newtonian sense, as
explained above) is also relevant for the exact relationship between the
angular deficit and the internal structure of the source, not only the
amount of matter as indicated in the relation (\ref{mi}).

The left hand side of equation (\ref{main}) does not depend on the cylinder
radius $\ell $. Thus, this relation is maintained even if we take the limit
$%
\ell \rightarrow 0,$ whenever the integrals on the right hand side converge
in this limit. Relation (\ref{main}) is also applicable to extended
distributions, as the complete string model described by the Lagrangian
(\ref
{lagrangian}). In this case, the junction condition must be considered
asymptotically at $\ell =\infty .$

For simple models in which $p_r=0$, relation (\ref{main}) reproduces the old
one (\ref{mi}). However, equation (\ref{main}) holds for more models than (%
\ref{exact}) does, since, the radial and hoop pressures do not have to be
zero in all interior points.

{\bf Acknowledgment.} I would like to thank Dr. Carlos Romero for
suggestions and for a careful reading of this paper. I am grateful to the
referees for their comments.


\begin{references}
\bibitem{vilenkin}  A. Vilenkin and E. P. S. Shellard, {\it Cosmic Strings
and Other Topologial Defects (}Cambridge U. P., Cambridge, 1994).

\bibitem{garfinkel}  D. Garfinkle, Phys. Rev. D {\bf 32}, 1323 (1985).

\bibitem{hiscock}  W. A. Hiscock, Phys. Rev. D {\bf 31}, 3288 (1985).

\bibitem{gott}  J. R. Gott, Ap. J. {\bf 288}, 422 (1985).

\bibitem{linet}  B. Linet, Gen. Rel Grav. {\bf 17}, 1109 (1985).

\bibitem{frolov}  V. P. Frolov, W. Israel and W. G. Unruh, Phys. Rev. D {\bf
%
39}, 1084 (1989).

\bibitem{geroch}  R. Geroch and J. Traschen, Phys. Rev. D {\bf 36}, 1017
(1987).

\bibitem{linet1}  B. Linet, Phys. Lett. {\bf A} {\bf 124}, 240 (1987).

\bibitem{futamase}  T. Futamase and D. Garfinkle, Phys. Rev. D {\bf 37},
2086 (1988).

\bibitem{hindmarsh}  M. Hindmarsh and A. Wray, Phys. Lett. {\bf B} {\bf
251}%
, 498 (1990).

\bibitem{israel}  W. Israel, Phys. Rev. D {\bf 15}, 935 (1976).

\bibitem{dahia}  F. Dahia and C. Romero, Mod. Phys. Lett {\bf A 14}, 1879
(1999).

\bibitem{moss}  I. Moss and S. Poletti, Phys. Lett. B {\bf 199}, 35 (1987).

\bibitem{peter}  P. Peter, Class. Quantum Grav. {\bf 11}, 131, (1994).

\bibitem{tolman}  R. C. Tolman, {\it Relativity, Thermodynamics and
Cosmology%
} ( Clarendon Press, Oxford, 1934).
\end{references}
\end{document}